# Technological improvement rate estimates for all technologies: Use of patent data and an extended domain description

Anuraag Singh[1*], Giorgio Triulzi[2,3] and Christopher L. Magee[3,4]

[1]Massachusetts Institute of Technology, Cambridge, MA, USA

[2]Universidad de los Andes, School of Management, Bogotá, Colombia

[3]Massachusetts Institute of Technology, Institute for Data, Systems, and Society, Cambridge, MA, USA

[4]SUTD-MIT International Design Center, Massachusetts Institute of Technology, Cambridge, MA, USA

*Corresponding Author

E-mail addresses: (AS) anuraag@mit.edu, (GT)  gtriulzi@mit.edu, (CLM) cmagee@mit.edu

## Abstract

In this work, we attempt to provide a comprehensive granular account of the pace of technological change. More specifically, we survey estimated yearly performance improvement rates for nearly all definable technologies for the first time. We do this by creating a correspondence of all patents within the US patent system to a set of technology domains. A technology domain is a body of patented inventions achieving the same technological function using the same knowledge and scientific principles. We obtain a set of 1757 domains using an extension of the previously defined classification overlap method (COM). These domains contain 97.14% of all patents within the entire US patent system. From the identified patent sets, we calculated the average centrality of the patents in each domain to estimate their improvement rates, following a methodology tested in prior work. The estimated improvement rates vary from a low of 1.9% per year for the *Mechanical Skin treatment- Hair Removal and wrinkles* domain to a high of 228.8% per year for the *Network management- client-server applications* domain. We developed a one-line descriptor identifying the technological function achieved and the underlying knowledge base for the largest 50, fastest 20 as well as slowest 20 of these domains, which cover more than forty percent of the patent system. In general, the rates of improvement were not a strong function of the patent set size and the fastest improving domains are predominantly software-based. We make available an online system that allows for automated searching for domains and improvement rates corresponding to any technology of interest to researchers, strategists and policy formulators.

## 1. Introduction

The pace of technological change has been discussed as a cause for rising anxiety and renewed concerns about social impacts of technological change (Jones, 2009; Gordon, 2012; Mokyr et al., 2015; Bloom et al., 2017; Groshen et al., 2019; O'Donovan, 2019; Autor, 2019). There have been academic studies of probability of job loss due to progress in specific technologies such as machine learning, mobile robotics among others (Frey and Osborne, 2017; Brynjolfsson and Mitchell, 2017; Brynjolfsson et al., 2018). In popular forums, technological change and disruption have been blamed for large scale layoffs by large multinationals (for instance- "The Weekly | G.M. Leaves Lordstown Behind in Hard Bet on Future," 2019). There have also been attempts to attribute significant political changes such as the US presidential elections to technological change, specifically advances in automation (Frey et al., 2017). Therefore, a granular yet broad and systematic understanding of technology and the pace of technological change appears to be a key step in enabling a more useful understanding of technological change and its societal effect.

However, most discussions of technological change are based around sector-specific changes. Moreover, such studies have often discussed pace of change in a qualitative way and have usually not defined the distinct boundary of the technology being studied. Despite the central importance of technology in driving economic growth (Schumpeter, 1935; Solow, 1957; Romer, 1990), there have been limited attempts at a systematic survey of the technological toolbox available to humanity that go beyond looking at a single technology classification system. In this paper, we build on previous work to describe 97.2% of the patent system as a set of 1757 discrete technology domains and quantitatively assess each domain for its improvement potential.

The rate of improvement of performance for a technology is an important indicator of the potential future importance of that technology (Hoisl et al., 2015). Consistent empirical evidence, accumulating since 1965, using datasets of performance time series for a variety of different technologies, shows that performance improvements for individual technologies follow exponential trends over time. This implies that technologies experience constant yearly rates of improvement, albeit having very different rates (Moore, 1965; Martino, 1971; Nordhaus, 1996; Moore, 2006; Koh and Magee, 2006; Nordhaus, 2007; Koh and Magee, 2008; Nagy et al., 2013; Koomey et al., 2011; Farmer and Lafond, 2016; Magee et al., 2016). Particularly significant is the work of Farmer and Lafond (2016) which rigorously shows that long-term trends are random walks around an exponential in time. Based on this body of prior work, we have reliable empirical measures for more than 30 technological domains. However, these measurements only cover less than 15% of the patents in the US patent system. Moreover, the rate of improvement can only be empirically estimated when substantial performance measurements are made over long time periods (Benson, 2014). In some large technological fields, including software and clinical medicine, such measures have rarely, if ever, been made. The major purpose of this paper is to provide estimates of the performance improvement rates for the hundreds of domains not accessed by empirical measurement. To our knowledge, this paper is the first to attempt completing a full-breadth yet granular survey of technological improvement rates. Prior related studies (Hoisl et al., 2015; Mokyr et al., 2015; Groshen et al., 2019; O'Donovan, 2019; Way et al., 2019) indicate that such results can be helpful in informing technological decisions for firms performing R&D as well as to inform policy makers trying to understand the social implications of technological change. In particular, to design industrial and social security policies and/ or prioritize R&D investments.

This work is based upon the prior empirical studies assessing performance improvement in 30 technologies defined as technological domains- sets of artifacts fulfilling a specific function using a specific branch of scientific knowledge (Magee et al., 2016). Existing work had demonstrated that patents corresponding to technological domains defined in this way can be reliably found using the classification overlap method (COM) described by Benson and Magee (2013, 2015a) and that such patent sets can be used to estimate improvement rates (Benson and Magee, 2015b). Recently, Triulzi et. al. (2018) have shown that accurate and reliable estimates of the rate of performance improvement can be obtained for technological domains based upon the average centrality of patents (in the patent citation network) in each domain. In this work, we use patent network centrality estimation methods and invert, automate & extend COM to the entire US patent system to provide estimates of performance improvement rates for the widest possible set of technologies. We also describe a new online interactive system where domains corresponding to technology-related keywords can be found along with their improvement rates[1]. The user can input a keyword describing the technology of interest through a user-friendly interface and the system returns an estimate of improvement for the technological domain, an automated measure of quality of match (called MPR) and patent sets so that the reader can judge the semantic quality of the match.

---

[1] The portal can be accessed at http://technologyrates.mit.edu/

## 2. Literature review and conceptual framework

In this section, we describe the major conceptual ideas underpinning this work- first, a technology domain description of technology; second, models of long-term technological performance improvement; third, retrieval of patents belonging to a particular technological domain using COM; and fourth, estimating rate of improvement using patent network information centrality.

### 2.1 Research into technological change has operated at three distinct levels-

On a detailed level, previous research, in particular in the field of technology and economic history, has usefully studied discrete technological inventions- such as Nelson's (1962) and Riordan and Hoddeson's (1999) study of the invention of the transistor, Enos's (1962) study of the series of inventions related to oil refining processes, the several studies on key inventions that characterized the first industrial revolution (Frenken and Nuvolari, 2004; Nuvolari, 2004) among others. Some have proposed lists of discrete inventions (such as Tushman and Anderson's list of technological discontinuities (1986) or the survey of significant innovations compiled at Science Policy Research Unit (SPRU) at the University of Sussex (Robson et al., 1988)). Others have analyzed prizes and awards for important inventions (Fontana et al., 2012; Capponi et al., 2019). There are however, difficulties in creating accurate and complete lists of inventions for tracking technological change. The process is labor intensive and requires considerable expert knowledge (Godin, 2007) and tends to overestimate the importance of singular inventions when much progress occurs through a series of inventions over time. Moreover**,** it is well-known that technologies have extensive interaction with one another (often called spillover) in that technological ideas can be used for various purposes and that prior technological and scientific ideas are at the root of even the most novel technologies (Usher, 1954; Koestler, 1961; Rosenberg, 1982; Dasgupta, 1996; Verspagen, 1997; Youn et al., 2015; Basnet and Magee, 2016).

On the broadest level, technology has been considered in neoclassical work as a single integrated unit or a black box (Solow, 1957). This macro perspective on the relationship between technology and growth has been complemented by a series of research efforts aiming at measuring productivity growth and productivity differentials across time, industries or countries (Brynjolfsson, 1993; Fagerberg, 1994, 2000; Gordon, 2000; Acemoglu and Robinson, 2010; Gordon, 2017). However, these efforts focused on measuring the effect of technological change, without necessarily explaining its sources and mechanism.

To do that, it is important to open the technological black box, as argued by Rosenberg,  (1982) and not treat technology as a monolith- *"Specific characteristics of certain technologies have ramifications for economic phenomena…"* such as productivity improvement, technological learning, technology transfer and effectiveness of government policies intended to influence technology. The body of work on sectoral (Malerba and Orsenigo, 1995; Breschi et al., 2000; Malerba, 2002), technological (Hekkert et al., 2007; Bergek et al., 2008; Markard and Truffer, 2008) and national (Lundvall, 1992; Nelson, 1993) innovation systems have expanded on this view by showing how these factors differ systematically across sectors and countries due, in great part, to the characteristics of the technological base.

Attempts to bridge the gap between discrete technologies and the black box have been an important aspect of research on technological change. Further advances in our understanding of the *process* of technological change came from its characterization as "progress", defined as the improvement of *"multi-dimensional trade-offs"*, along precise technological trajectories (Dosi, 1982) and by the development of methods to map and analyze these trajectories empirically (Verspagen, 2007; Castaldi et al., 2009; Nuvolari and Verspagen, 2009). A crucial advance in the field, came from the development of precise definition of technology and the diffusion of a shared understanding of it. In his seminal work, Dosi (1982) defines technology as combinations of "theoretical" and "practical" know-how, methods, processes, experiences as well as their embodiments in physical devices and equipment. Arthur adds a complementary perspective and defines technology as *"a means to fulfill a purpose, and it does this by exploiting some effect"* in his work, the structure of invention (Arthur, 2007). Magee et al. (2016) have built on prior work by Dosi and Arthur to define Technological Domain (TD) as *"The set of artifacts that fulfill a specific generic function utilizing a particular, recognizable body of knowledge."* This definition introduces TD as a means to avoid the confusion associated with the word technology which has come to mean widely different things to different people. Magee et al. (2016) have further employed the concept of technological domains to obtain reliable empirical estimates of technology improvement rates for 30 domains over periods of decades.

In sum, the field of research of technological change now has precise definitions of what technology is, how it evolves, why it does it in ways that differs across sectors and countries and how they affect economic growth differentials at these two levels. Nevertheless, despite these extensive efforts to study what determines the *direction* of technological change and how it evolved in selected industries, the *rate* of improvement of technological advances and its differences across technologies, a key determinant of economic growth *differentials*, has not been studied convincingly and systematically. Our work aims at contributing to filling this gap.

**2.2 Long term performance improvements in technologies can be modeled as exponential change**

Performance over time, of components or product generations, has sometimes been characterized by S-curves (Utterback and Abernathy, 1975; Sahal, 1981; Tushman and Anderson, 1986; Christensen, 1992a, 1992b; Ayres, 1994; Christensen, 1997; Schilling and Esmundo, 2009). However, in his seminal paper Christensen (1992b) points out that the flattening part of the S-curve for individual components in his data is a firm-specific phenomenon. Similarly, Henderson (1995) saw the flattening in performance of optical photolithographic alignment technology as not persistent and thus, non-existent in the long term. Most of the other studies quoted are not long-term. However, long-term performance improvement rates, defined as the *"…trend of non-dominated (i.e. record-breaker) performance data points for the overall technology domain (not for individual product generations, individual companies or components)"* (Triulzi et al., 2018), are critical for long-term strategy and technology management.

As discussed in section 1, a consistent body of empirical evidence shows that performance improvements for individual technologies follow exponential trends over the long-term, consistent with constant yearly rates of improvement (Moore, 1965; Martino, 1971; Nordhaus, 1996; Moore, 2006; Koh and Magee, 2006; Nordhaus, 2007; Koh and Magee, 2008; Koomey et al., 2011; Nagy et al., 2013; Farmer and Lafond, 2016; Magee et al., 2016). Studies of large sets of such data (Farmer and Lafond, 2016; Magee et al., 2016) agree that random walk around the exponential (constant yearly % increase) is the most appropriate description. Short term segments of these noisy exponentials can be described as S curves but these do not hold up in the long-term as evident in the Farmer and Lafond analysis (2016). Moore's law (1965) is the single most famous example of exponential long-term technology improvement. Thus, we refer to the fact that all domains show this exponential behavior as the Generalized Moore's Law (GML).

The performance of many technologies over time $Q_i(t)$ can then be expressed by the following mathematical description given in Equation 1. $Q_i$ represents the intensive performance metric, subscript $i$ denoting technological domain $i$ and subscript 0 denoting time equals $t_0$.

$$Q_i(t) = Q_{i0} exp\{k_i(t - t_0)\} \tag{1}$$

The exponential factor ($k_i$) in Equation 1 is domain dependent. While different technologies all improve exponentially, they do so at different rates (Koh and Magee, 2008, 2006; Magee et al., 2016). However, $k_i$ is constant (at least to a good approximation) over time in a domain (Farmer and Lafond, 2016) and for different productivity metrics within a domain (Magee et al., 2016).

The characterization of innovation as a combinatoric process of existing ideas (Usher, 1954; Ruttan, 2000; Fleming, 2001; Fleming and Sorenson, 2001; Frenken and Nuvolari, 2004; Frenken, 2006a, 2006b, 2006c; Weisberg, 2006; Gruber et al., 2012; Youn et al., 2015) has been suggested to explain why exponential improvements are observed (Youn et al., 2015; Basnet and Magee, 2016). Furthermore, fundamental properties of a technology domain, such as scaling laws and the complexity of interactions between the components artifacts, have been conjectured to determine differences in rate of performance improvement across domains (Dutton and Thomas, 1984; McNerney et al., 2011; Basnet and Magee, 2016). Thus, while the form of Equation 1 seems to suggest that performance depends only on an exogenous time-trend, it captures improvements due to science, spillover from other technologies, scaling (increase in production) and complexity of interactions (modularity).

A different description for decrease in cost and increase in performance, is based on the observation that cost of many technologies decreases as a power law with cumulative production. The phenomenon is known as an experience curve, learning curve, or Wright's law suggesting learning-by-doing processes as the possible cause (Wright, 1936; Argote and Epple, 1990). Ayers and Martinàs (1992), taking a more expansive view, argue that the experience curve is not just related to learning-by-doing but an indirect measure of total effort, including *"… incremental design improvements, increased capital intensity in the manufacturing process and (closely related) economies of increasing scale"*. However, experience curves and their conceptualization as examples of learning-by-doing have been subject to criticisms. Nordhaus (2014) has shown that there is a fundamental statistical problem in separating learning processes from exogenous technological change during modeling and attributing causality. Sinclair et al. (2000) as well as Funk and Magee (2015) have shown the important role of R&D which is missed in simple models based on experience curves. Magee et. al (2016) examined the relationship of the number of patents over time with technical performance and found that Moore's Law holds even when the number of patents do not increase exponentially with time. This suggests that Moore's Law is fundamental over the long-term and independent of "effort" variables such as total number of patents[2].

The empirical evidence on Moore's and Wright's law and the apparent difference in interpretation, can be reconciled by the mathematical fact that the two laws are equivalent as long as cumulative production increases exponentially over time (Sahal, 1979; Nagy et al., 2013; Magee et al., 2016). More recently, Lafond et al. (2020) have shown that growth of experience (measured by cumulative production) and an exogenous time trend (excluding cumulative production) contributed roughly equally to the decreases in cost of military products during world war 2.

---

[2]Past studies have shown that number of patents are significantly correlated with the research "effort" in the domain measured by investment- specifically R&D spending. See Griliches, Z., 1990. Patent Statistics as Economic Indicators: A Survey (Working Paper No. 3301). National Bureau of Economic Research. https://doi.org/10.3386/w3301 for an excellent review.

**2.3 Patents represent the core technology invention and the Classification Overlap Method (COM) makes patent retrieval repeatable**

Patents are a set of data that contains the raw information created by the inventors of millions of patents over hundreds of years, and additionally by input from thousands of expert patent examiners whose knowledge is embedded in the organization of this massive data set. Another positive attribute of patents is that they are focused on the activity (invention) that is the mediator of the two other key activities in technological change (pursuit of scientific knowledge and product development). However, many technological progress researchers find the categories defined by the patent examiners "too detailed" and inadequate in representing the reality of the technological enterprise (Larkey, 1999; Hall et al., 2001). For this work we use a decomposition based on technology domains described in Section 2.1.

The patents corresponding to technological domains defined above can be reliably found using the classification overlap method (COM) described by Benson and Magee (2013, 2015a). COM is an improvement over the traditional keyword search and the classification search and makes patent retrieval repeatable. The usage of two separate hierarchical classification systems by the USPTO (up to mid 2015) allowed distinction between function and knowledge base (the two basic concepts underlying technological domains) to be built into the classification scheme. The resulting success of COM in retrieving patents that are consistent with the artifacts whose performance improvement is measured in the domains is the fundamental reason for expanding the coverage to previously unidentified domains in the current work. Operationally, the normal use of COM first retrieves all patents using a pre-search based set of keywords (in the patent title or abstract), companies or individual inventors. The most representative technology classes belonging to both International Patent Classification (IPC) and the United States Patent Classification (UPC), are ranked using an objective score. It is important to note that many patents are classified into multiple classes (Benson and Magee, 2015a; Magee et al., 2016) and so a patent might appear multiple times in each of the most representative technologies. Finally, all patents that have been classified in both the topmost representative IPC class and the topmost UPC class are retrieved for that domain.

In the current research, we invert this approach by examining all possible domains- all possible IPC and UPC class overlaps studying all that have statistically significant numbers of patents. The patents found this way are coherent and in understandable technological areas, allowing us to discern the function, context and evolution of a domain. More importantly, they can also help us arrive at quantitative estimates of technological improvements.

**2.4 Estimates of the rate of performance improvement can be made by using patent network information centrality of patents belonging to a domain**

As shown in Benson and Magee (2015b) and, more recently, by Triulzi et al. (2018), once a patent set for a technology domain has been identified, it is possible to estimate the yearly rate of performance improvement for that domain. In these two papers the authors tested several different patent-based measures as predictors of the yearly performance improvement rate for 30 different technologies for which observed performance time series were available. By far, the most accurate and reliable indicator is a measure of the centrality of a technology's patents in the overall US patent citation network, as shown in Triulzi et al. (2018). More precisely, technologies whose patents cite very central patents tend to also have faster improvement rates, possibly as a result of enjoying more spillovers from advances in other technologies and/or because of a wider use of fast improving technologies by other technologies, proxied by patent citations. The measure of patent centrality used is a normalized version of the "Search Path Node Pair" (SPNP) index proposed by Hummon and Doreian (1989) and operationalized in a fast algorithm by Batagelj (2003) for directly acyclical graphs and popularized by, among others, Verspagen (2007) to identify the main paths of technological development in a patent citation network. The SPNP index is a measure of information centrality, conceptually similar to the random-walk betweenness centrality. It measures how often a given node shows up on any path of any length connecting any given pairs of nodes in the network. Therefore, central patents are like information hubs in the citation network, representing inventions that are related technologically by a path of improvements to many other inventions that appeared before and after them.

Triulzi and colleagues, normalized the centrality index by randomizing the citation network under a set of constraints, such as the indegree and outdegree of each patent, the share of citations made by each patent that goes to the same main technology field of the focal patent and the age of the citing-cited pair for each citation (for more information see Appendix B). This makes centrality comparable for patents granted in different moments in time and assigned to different technology fields, which, in turn, allows computing a comparable average centrality for patents across technology domains. The latter was shown to have a correlation of 0.8 with the log of the yearly improvement rate. As a result, the authors showed how the following estimated equation, trained by running a regression for the 30 technologies for which observed improvement rates are available, can be used for out-of-sample predictions of the improvement rate of any given technology domain i for which an accurate patent set can be identified[3].

$$Estimated\ K_i = e^{(6.15987*X_i - 5.01885)} * e^{\frac{\sigma_i^2}{2}} \tag{2}$$

---

[3] For a discussion of the prediction intervals of the estimation we refer to Triulzi et al. (2018).

In Equation 2, numbers inside the bracket are the estimated coefficients of an OLS regression that has the log of the improvement rate as dependent variable, an intercept and one predictor $X_i$ for each technology domain $i$. In Triulzi et al. (2018), this predictor is the mean value for all patents in domain $i$, of the average centrality of the patents cited by each patent $j$ in domain $i$. The second term in the right-hand side is a correction factor to move back from a log scale to a linear scale.

## 2.5 Overall Framework

We build on the concepts described above to, first, decompose the entire patent system into a set of technology domains by extending, inverting and automating COM. Second, we calculate rates of improvement for each of the domains belonging to the above set. Third, we identify some of the key technological domains. Fourth, we provide a new online system for searching technologies and their improvement rates.

**3. Data**

Our dataset contains all patents issued by USPTO from 1976-2015 for which valid U.S. Patent Classification system (UPC) and International Patent Classification (IPC) current classification data exist. We used the current classification data files i.e. reclassified data and not the data at time of grant. We also use the list of 3-digit current UPC classes and 4-character IPC classes for the extension of COM described in section 4.

We obtained the dataset, with all patents granted since 1976, from the PatentsView platform[4]. PatentsView gets access to the data through an arrangement with the Office of Chief Economist in the US Patent and Trademark Office[5] and is current through October 8, 2019. The dataset contains patent number, date of grant and other metadata. We limit our dataset only to the US patents because the performance datasets available to us are overwhelmingly from the US and because the UPC system is necessary for application of COM. We do believe that US patent data is representative of patenting activity worldwide due to its reputation as a technology leader and the vast size of the consumer market enticing most global firms to patent in US.

We only consider patents with grant dates between 01-01-1976 to 06-01-2015 totaling 5.7 million. We remove non-utility (special) classes of patents such as those with the designation "D", "PP", "H", "RE" and "T"[6], summarized in Table 1. This yields a total of 5083263 valid unique utility patents.

| Patent Type | Number of patents |
|:---:|:---|
| D-type | 573505 |
| PP-type | 25221 |
| H-type | 2258 |
| RE-type | 17955 |
| T-type | 509 |

*Table 1 Number of patents eliminated from the patent set*

The USPTO updates the taxonomy at regular intervals to maintain 'consistentcy' in classification (in addition to ease of searching) as the meaning of 'consistent' changes over time (Lafond and Kim, 2017). We use the complete list of 3-digit current UPC classes (439 in number, obtained from the USPTO

---

website[7]) and 4-character IPC classes (648 in number, obtained from the WIPO website[8]) for utility patents. We exclude the UPC class "G9B" because of its very high similarity to the corresponding IPC class from which it originated, thus, rendering it in unsuitable for COM. The UPC classes list has not been updated since May 2015. The IPC classes list continues to be updated every year and we used version 2019.1.

The USPTO also reclassifies patents so that, the patents adhere to the lastest taxonomy. We use the reclassification data as we believe that the current structure of technology is best reflected in the patent classification we have now, instead of the one at the time of grant. Using classification at time of grant we arrive at a slightly different set of domains (both number and composition). For a historical analysis we would use both reclassified data and the classification at time of grants to understand the evolution of structure of technological domains but such an analysis is beyond the scope of the current paper.

The UPC current classification data were downloaded from the PatentsView platform[9]. The classification data is based on USPTO bulk data files which were last updated on 2018-05-18[10]. There are 22,880,877 patent records with the current UPC classification data. These contain 5134285 unique patents suggesting each patent belongs to 4.46 UPC classes.

The IPC current classification data was obtained from the Google BigQuery platform[11] which uses data from IFI CLAIMS Patent Services. The UPC to IPC concordance was last published in 08/20/2015[12]. As such, no reclassification data for IPC is available after 2015. There are 21,857,265 patent records with International Patent Classification system (IPC) classification data (from 1976 to 2019). These contain 5920113 unique patents suggesting each patent belongs to 3.69 IPC classes. As noted in section 2.3, we use all classes in which a patent (both UPC and IPC) is listed and not just the main class.

---

[7] https://www.uspto.gov/web/patents/classification/selectnumwithtitle.htm
[8] https://www.wipo.int/publications/en/details.jsp?id=4442&plang=EN
[9] https://www.patentsview.org/download/
[10] https://bulkdata.uspto.gov/data/patent/classification/
[11] https://bigquery.cloud.google.com/dataset/patents-public-data:patents
[12] https://patents.reedtech.com/classdata.php

## 4. Methodology

This section provides details on the methods employed in each of the four steps outlined at the end of section 2 and describes how we use the data from section 3.

### 4.1 Decomposition of the entire patent system into a set of technology domains

We build on the concept of domains and the discovery of patents belonging to a particular domain using the classification overlap method (COM) described in Section 2 and describe the extension, inversion and automation of COM to give a technology domain description for the entire patent system. We don't start with a pre-search for a technology of interest as in the usual COM application. Instead, we start with the set of patents described above as well the lists of UPC and IPC classes. We pick one class from the UPC list and one from the IPC list and find all patents which belong to each of those classes using the classification data. We then find the "overlap" between these two sets- the patents which lie in both the given IPC class as well as the given UPC class. We do this for all possible class pairs i.e. unique combinations of classes- one from IPC and one from UPC and thus define the full set of overlaps. All overlaps are potentially domains but only if a large enough set of patents occupies the overlap.

To illustrate this, we show the overlaps between UPC classes 850, 353 & 123 and IPC classes G01Q, F02B & H02B in Table .

| Class No. | Class Name | Class Size |
|---|---|---|
| **850** | Scanning-probe techniques or apparatus | 3045 |
| **353** | Optics: image projectors | 9282 |
| **123** | Internal-combustion engines | 62113 |
| **G01Q** | Scanning-probe techniques or apparatus; applications of scanning-probe techniques | 4748 |
| **H02B** | Boards, substations, or switching arrangements for the supply or distribution of electric power | 5147 |
| **F02B** | Internal-combustion piston engines; combustion engines in general | 35318 |

*Table 2 Illustrative list of 3 UPC and 3 IPC classes each with their number, name and size*

This yields a total of 9 potential domains as shown in Table where the values in each intersection are the number of patents found in that overlap. For instance, for the class pair 123F02B there are 20,575 patents that are listed both in the 62,113 patent UPC 123 class and the 35,318 patent IPC F02B class.

| | | IPC | | |
|---|---|---|---|---|
| | Class No. | G01Q | H02B | F02B |
| **UPC** | **850** | 2946 | 0 | 0 |
| | **353** | 0 | 2 | 0 |
| | **123** | 0 | 0 | 20575 |

*Table 3 Illustration of a subset of overlaps. UPC Classes are listed row-wise and IPC classes are listed column-wise. We then find the overlap between them finding all patents which lie in both the UPC as well as the IPC class.*

We systematically, calculate the overlap for all possible class pairs. Since each class pair is composed of an IPC class and a UPC class, there are 284472 possible class pairs in total. We obtain the overlap as the size of intersection of the set of patents in the selected UPC class and the set of patents in the selected IPC class (i.e. the set of patents listed in both the UPC and IPC class being combined). We find that most of these overlaps are empty- 55% of the overlaps are zeros.

We only consider those class pairs as domains which have above random probability of being in an overlap. This is done to avoid misclassification noise (below we will see this eliminates the domain label from class pair 353H02B containing only 2 patents). To deduplicate the patent sets, we then assign patents which lie in more than one overlap to the biggest overlap that they occupy. We thus obtain a final list of domains with each patent matched to only a single domain. Each of these steps is described in more detail below.

To assess potential mis-classification and/or typos as a source of noise, we calculate the expected probability of patents lying in a overlap given as the product of probability that a patent lies in a given IPC class x- $P(IPC\_x)$ and the probability that the patent lies in the given UPC class y- $P(UPC\_y)$, if they were independent events. If the UPC and IPC patent classes are unrelated i.e. the probability of being classified in the given IPC class is independent of being classified in a given UPC class, then the joint probability $P(IPC\_x \cap UPC\_y)$ is in principle the probability of <u>randomly</u> misclassifying due to a typing mistake (typo) or a thinking mistake (thinko). In general, in a domain, the IPC class and the UPC class should be more than randomly related. Therefore, if the overlap is less than that of the patent being randomly classified in both the given UPC class and the IPC class, we discard that overlap as it is not an actual domain. We now show a worked example.

The probability of any joint event $A$ **and** $B$ i.e. $P(A \cap B)$ equals the product of probability of event $A$ i.e. $P(A)$ and probability of event $B$ i.e. $P(B)$ if the two events are completely independent-

$$P(A \cap B) = P(A) . P(B) \tag{3}$$

Given the value of $P(IPC\_x \cap UPC\_y)$ and the size of the sample space (total number of patents in our set), we can obtain the expected size of the overlap. For instance, in Table 3, for the class pair 353H02B we calculate:

$$P(UPC\_353) = number\ of\ patents\ in\ UPC\ class\ 353/\ Total\ number\ of\ US\ patents \tag{4}$$

Thus, $P(UPC\_353)$ = 9282/5083263 = 0.0018. Similarly, $P(IPC\_H02B)$ = 0.001.

The joint probability,

$$P(UPC\_353 \cap IPC\_H02B) = P(UPC\_353)\ x\ P(IPC\_H02B) = 1.85\ x\ 10^{-6} \tag{5}$$

Finally, the expected overlap can be calculated as:

$$Expected\ overlap = P(UPC\_353 \cap IPC\_H02B)\ x\ Total\ number\ of\ US\ patents \tag{6}$$

The randomly expected overlap comes out to be 9.4. The actual overlap is only 2. Thus, we regard this class pair as not being a domain and do not analyze it further. We discard all class pairs with actual overlap less than randomly expected to indicate that the overlap could occur because of noise due to miswritten class numbers or other semi-random noise. In all, we lose 23711 patents accounting for 0.47% of total patents and finally, obtain a set of valid "domains". For efficiency, we also, discard all class pairs which contain less than 100 patents as we believe that is a reasonable threshold for a set of patents to constitute a technology domain with a coherent function and knowledge base.

Since some patents lie in multiple UPC and multiple IPC classes (as discussed in Section 2.2), some patents naturally lie in more than overlap. For simplicity and ease, we assign them to the largest overlap so that the final decomposition lists each patent in only one domain. For the purposes of technology improvement rate this does not make a big enough difference (see Appendix A) to concern us as this work is focused on rate of improvement. In research on technological structure, the duplicated lists would be used as well but this is beyond the scope of the current work. Deduplication empties a number of small overlaps and reduces the number of patents in others. Going forward, by size we mean the number of unique patents after deduplication. With reference to our illustrative example, this results in 850G01Q and 123F02B as valid domains with sizes 568 and 20437 (smaller than the original overlaps due to deduplication) as shown in Table 4.

| | | IPC | | |
|---|---|---|---|---|
| | Class No. | G01Q | H02B | F02B |
| **UPC** | **850** | **568** | 0 | 0 |
| | **353** | 0 | discard | 0 |
| | **123** | 0 | 0 | **20437** |

*Table 4 Overlaps from Table 3 after deduplication and elimination of noise. The numbers in red show class pairs which become domains with their size after deduplication. Grey depicts discarded and empty pairs.*

Figure 1 shows a graphical summary of the methodology and the key steps described above-

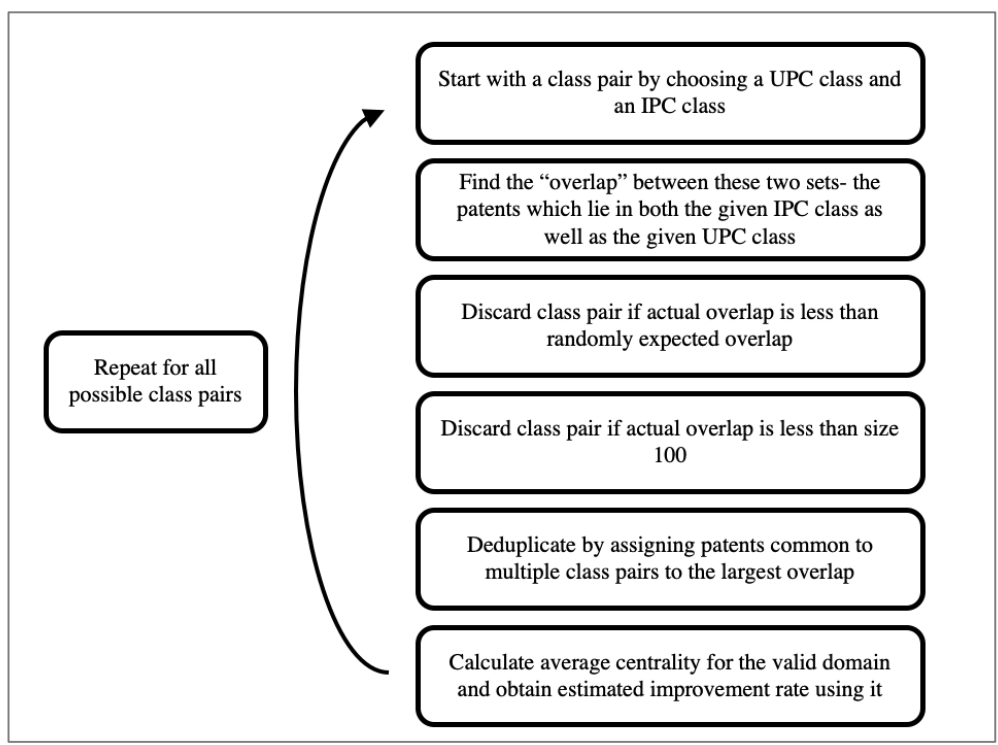

*Figure 1 Process steps for inverted COM*

Recall that there are 283824 (438 x 648) potential domains having a unique UPC and IPC classification and 5083263 utility patents in our set that have appropriate UPC and IPC designations. **66.3% i.e. two-thirds of these patents** are contained in the largest **175 domains** which is only **0.06% of the possible domains** showing that technologies are selectively in a relatively narrow set of domains. Indeed, **13142** overlaps (i.e. 4.62% of class pairs) contain 99.52% of all patents[13]. The details are shown in Table 5.

| Size of Domain | Number of domains of that size | Number of domains of that size as a fraction of total possible | Total number of unique patents in all domains of that size | Fraction of unique patents in all domains of that size |
|---|---|---|---|---|
| **1-9** | 8400 | 0.0296 | 23864 | 0.0047 |
| **10-99** | 2985 | 0.0105 | 94163 | 0.0185 |
| **100-999** | 1153 | 0.0041 | 390087 | 0.0767 |
| **1000-9999** | 490 | 0.0017 | 1685323 | 0.3315 |
| **Above 10000** | 114 | 0.0004 | 2865248 | 0.5637 |

*Table 5 Domain size distribution and patent coverage as a function of domain size.*

Table 5 illustrates two results. First, most of the domains are small domains (86.6% of the domains contain less than 100 patents each) and secondly that despite the large number of small domains, most patents are in larger domains (almost 90% of the patents are in domains that have at least 1000 patents). For our later survey of rates of improvement, we confine the domains considered to those containing greater than or equal to 100 patents. As can be found by adding the last three entries in Table 5, this yields 1757 domains and 97.14% of the patent system.

---

[13] Recall that before deduplication, only 55% of the possible domains contained zero patents. Indeed, the overall concentration is stronger after deduplication.

**4.2 Calculation of rates of improvement for the set of domains**

As explained in Section 2.4, we use the method defined in Triulzi et al. (2018) to estimate the improvement rate for each identified technology domain. However, we depart from that paper in the choice of which centrality measure to use. In Triulzi et al. (2018), the authors propose using the average normalized centrality of the patents cited by a domain's patents as a predictor of the domain's improvement rate. This is done because using data on the focal patents' centrality would require waiting an arbitrary number of years after the patent is granted to allow the patent time to accumulate citations, which is necessary to measure its centrality reliably. Since a focal patent's centrality and the centrality of the patent it cites are strongly correlated and given that in Triulzi et al. (2018) some of the domains studied were very recent, in that paper the authors preferred to use the centrality of the cited patents to avoid losing data for the young domains. However, in this work, we are analyzing a very large sample of domains, most of which are fairly old. Therefore, we prefer using the normalized centrality of the focal patents in a domain computed after three years from the moment the patent is granted, given its stronger appeal in terms of ease of computation and presentation of the measure. We use the following equation, adapted from Triulzi et al. (2018).

$$Estimated\ K_i = e^{(6.217219 * X_i - 4.974221)} * e^{\frac{\sigma_i^2}{2}} \tag{7}$$

The coefficients have been obtained by training an OLS regression of the log of the observed improvement rate for 30 technologies (for which empirical time series of performance over time were available) against the average normalized centrality of their patents measured three years after being granted ($X_i$ in the equation). The improvement rates for these 30 technologies and their patent sets with centrality values are the same used in Triulzi et al. (2018).

For each of the 5083263 utility patents granted by the USPTO between 1976 and 2015, we compute the normalized centrality index using the same citation network randomization procedure presented in Triulzi et al. (2018) and explained briefly in Appendix B. We then compute the average centrality of patents in each of the 1757 identified technology domains and plug it in the equation to obtain the estimated yearly performance improvement rate.

It is important to note that the normalization of the centrality measure for each patent granted by the USPTO, produced an indicator that is uniformly distributed between 0 and 1. Therefore, if patent sets for the technology domains are sampled randomly from the overall set and then the average normalized patent centrality for each technology domain is calculated, its distribution would follow a normal distribution with mean equal to 0.5. If that were to be true, the distribution of the estimated improvement rate could not be interpreted as it would just be an artifact of random sampling patents and of the centrality normalization method. This is not the case as all normality tests for the distribution of mean centrality across domains reject the normality hypothesis. In fact, the best fit for this distribution is an exponentially modified Gaussian (the sum of an exponential random variable and a Gaussian one). In Appendix C, we report details for the normality tests and the best fit. This methodological result, along with the test for randomly expected overlap (section 4.1), further strengthens our belief that our method is revealing an underlying property of technology system concerning the distribution of the improvement rates and patent centrality across domains. In Section 5.1, we show the distribution of the estimated improvement rate across the 1757 domains.

**4.3 Identification of domain technology**

The 50 biggest, 20 smallest and 20 fastest domains were studied in more depth. Each of them was then given one-line descriptions by two of the authors working separately and independently. Author 1 relied on meta data from the patent classification systems (class titles and definitions), similarity with older named domains (Benson and Magee, 2015a, 2016) and the description of top 20 central and randomly sampled 20 patents. Author 2 mostly relied on titles and abstracts of top 20 central and randomly sampled 20 patents from each domain using the popular NumPy library in Python programing language. In about 90% of the cases, agreement on the names/descriptions was easily seen and in the remaining 10% of cases, further joint work eliminated the discrepancy usually through more careful joint examination of sample patents.

**4.4 Online technology search system**

The results presented in Section 5 provide a broad and systematic account of technological change. However, improvement rates for specific technologies (or domains) or groups of related technologies are also of potential interest to many researchers and policy makers. Thus, we have developed an online technology search system which enables a user to find estimated improvement rates for a technology of interest.

For each given search term, we search patent title and abstracts across the entire dataset of valid US utility patents (with grant dates between 01-01-1976 to 06-01-2015) and return the list of patent numbers containing the term. This is accomplished by using full text search functionality in a relational database (such as MySQL, PostgreSQL etc.). The standard text search function incorporates tokenization, stemming and vectorization to enhance the search. We then match this list of patents to our corresponding domains by using the correspondence established before. We find the most representative domain for those patents by using a relevance ranking. The relevance ranking for the patent classes is accomplished by using the mean-precision-recall (MPR) value proposed by Benson and Magee (2013). This value was inspired by the 'F1' score that is common in information retrieval, but uses the arithmetic mean (instead of the geometric mean) of the precision and recall of a returned data set (Magdy and Jones, 2010). We return the top 5 most representative domains along with an estimate of the improvement rate for each domain as well as the title and abstract of the top 20 and a random set of 20 patents from the most representative domain.

The user is then able to judge whether to try different key words if the example patents indicate something different than what they intended to examine or want to pursue interesting leads from reading the patents they discover in the first round. The search tool can be accessed by the readers from the project website[14] through a user-friendly interface and is hosted on a cloud server.

---

[14] The portal can be accessed at http://technologyrates.mit.edu/

## 5. Results and Discussion

Using the extended, inverted and automated COM and K-estimation methods defined above we were able to obtain estimated improvement rates for 1757 technologies which constitute 97.14% of the overall patent system. The key results are now presented.

### 5.1 Most technologies improve moderately or slowly-

Given that this is the widest collection of technologies and their improvement rates, we feel we can answer with some confidence the question- at what rates do technologies improve? On average- moderately or slowly.

More than 82.7% of the technological domains are improving at a rate of less than 25% per annum and more than 60% at a rate less than 12.5%. Performance for the average technology improves at a rate of 19.19 % per annum, with a standard deviation of 0.2625 (i.e. 26.25%). Figure 2 shows the distribution of the estimated improvement rate values for the 1757 domains and the best fit for the probability density function. The distribution is very skewed and its best fit is log-normal with shape, location and scale parameters respectively equal to 0.9424, 0.018 and 0.1014[15]. Note that the histogram resembles Figure 3 from Farmer and Lafond (2016) which plots the distribution of improvement rates for empirical data for cost improvement relating to about 53 technologies. This provides some additional evidence that the quality and reliability of our estimating method is adequate to provide meaningful estimates.

---

[15] The distribution has been fit with the Python package Fitter (https://fitter.readthedocs.io/en/latest/). The log-normal distribution parameter interpretation is reported in the Python SciPy Stats package (used by Fitter) here: https://docs.scipy.org/doc/scipy/reference/generated/scipy.stats.lognorm.html.

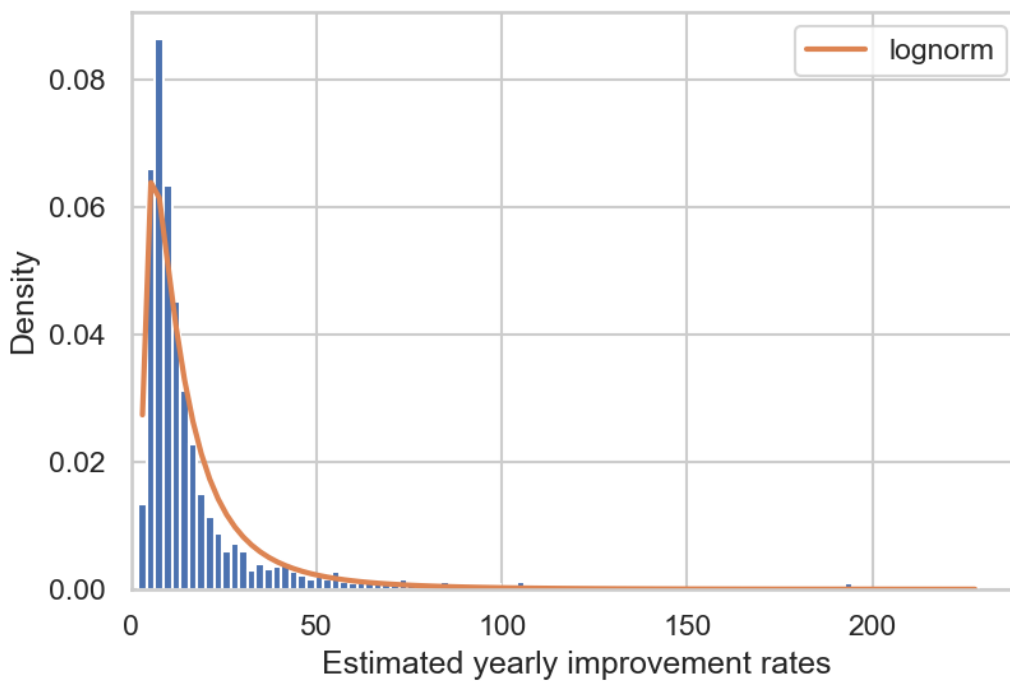

*Figure 2 Distribution of Estimated Improvement rates (K-values) for 1757 technology domains and best fit of the probability density function*

The identification of the technology in the largest 50 domains is given in Table 6 (by the method explained in section 3.3). The domains are identified by the Domain code- a portmanteau of the UPC code and IPC code of the underlying overlapped classes. The estimated improvement rate (estimated K) i.e. the values from Equation 2 are also given along with the domain size i.e. the number of patents (after deduplication). The estimated rates of improvement are reported as percentage change per annum. The improvement rates vary from a low of 5.6% per year for *Enzymatic reactions for synthesis/ detection/ amplification of biological molecules* domain to a high of 228.8% per year for the *Network management-client-server applications* (709G06F) domain. We note that the slowest of these 50 domains is the sixth largest while the fastest is the fifth largest anticipating the small relationship between size and rate of performance increase that will be more generally seen in Section 5.4.

**Table 6 Biggest 50 domains**

| Domain Code (UPC-IPC) | Domain Size | Estimated K | Domain Description |
|---|---|---|---|
| 257H01L | 233828 | 42.6 | Semiconductor devices and their fabrication |
| 514A61K | 118270 | 7.1 | Bio-affecting pharmaceuticals |
| 348H04N | 83929 | 46.7 | Display and Transmission of Digital Videos |
| 370H04L | 73936 | 113.9 | Routing and switching in a packet-switched computer network |
| 709G06F | 64666 | 228.8 | Network management specifically client-server applications |
| 435C12N | 59221 | 5.6 | Enzymatic reactions for synthesis/ detection/ amplification of biological molecules |
| 707G06F | 53644 | 178.1 | Data management (including databases and novel data structures) for enabling and automating ecommerce activities |
| 424A61K | 52436 | 8.9 | Surface active medicines |
| 439H01R | 51517 | 17.0 | Electrical connectors |
| 600A61B | 49611 | 23.1 | Medical Diagnostics- surgical, IV and non-invasive |
| 365G11C | 48935 | 41.2 | Solid state memories |
| 359G02B | 45698 | 19.6 | Optical elements for image generation, manipulation and enhancement recently emphasizing electro-optics |
| 455H04B | 42578 | 38.8 | Radio-frequency (Rf) management in radio-based communication networks |
| 382G06K | 41081 | 42.1 | Digital Watermarking and image analysis |
| 347B41J | 40738 | 24.1 | Non-impact, non-xerographic printing |
| 360G11B | 40651 | 19.5 | Magnetic Information Storage |
| 399G03G | 38257 | 23.1 | Xerographic printing with special emphasis on paper handling (Duplex printing) |
| 606A61B | 37599 | 19.0 | Electrosurgery |
| 345G09G | 36767 | 60.9 | Control of display devices |
| 428B32B | 36210 | 11.0 | Multilayered structures including coating and laminates for heat and wear resistance |
| 711G06F | 35418 | 109.0 | Access management for memory systems including caching |
| 324G01R | 35244 | 21.0 | Measurement of Electrical and magnetic fields |
| 705G06Q | 33997 | 122.8 | eCommerce including provision of financial and healthcare services |
| 604A61M | 31067 | 20.8 | Targeted drug/fluid delivery thorough catheter, stents etc. |
| 429H01M | 30887 | 13.3 | Non-aqueous batteries and fuel cells |
| 385G02B | 29457 | 44.9 | Fiber optic transmission |
| 528C08G | 29384 | 12.1 | Synthetic resins exhibiting superior thermal, solubility, electrical and mechanical properties |
| 264B29C | 28702 | 11.4 | Injection moulding and Heat treatment of plastics |
| 714G06F | 27584 | 99.5 | Digital circuit diagnostics and fault tolerant systems |
| 210B01D | 26936 | 10.3 | Fluid separation through filtration, adsorption, chromatography, reverse osmosis etc. |
| 455H04W | 26863 | 109.9 | Enabling mobility in Wireless networks |
| 713G06F | 26521 | 121.0 | Data safety in distributed computing environments |

| 379H04M | 26418 | 70.5 | Data-driven applications on Phones particularly VOIP, remote control of devices and messaging etc. |
|---|---|---|---|
| 369G11B | 24832 | 45.6 | Optical Information Storage |
| 166E21B | 24810 | 6.5 | Oil and gas well drilling operations including improving efficiency |
| 73G01N | 23255 | 13.3 | Material measurement and testing |
| 349G02F | 22378 | 128.6 | System design of LCDs |
| 375H04L | 22375 | 35.0 | Signal Processing for transmission loss prevention/ compensation in multichannel digital communication networks |
| 340G08B | 21836 | 25.7 | Electronic Security systems for physical security |
| 362F21V | 21777 | 16.6 | Lighting system and fixtures addressing reflective design, cooling etc. for various specific purposes of lighting system |
| 361H05K | 21025 | 20.6 | Containers and casings for electronic devices to enable thermal dissipation, effective EMR shielding, waterproofing etc., including production of flexible wearable electronics |
| 715G06F | 20979 | 142.1 | Compilation of related content/actions into a user-friendly graphic user interface based on context |
| 310H02K | 20871 | 8.8 | Structural improvements in electric motors and engines |
| 526C08F | 20664 | 10.6 | Process improvement in unsaturated carbon bond polymerization reactions |
| 123F02B | 20437 | 14.0 | Internal combustion engines- subsystems design and system architecture variants |
| 206B65D | 20418 | 7.3 | Packaging and Containers providing protection against physical harm, leakage, spillage, thermal spoilage, light-spoilage |
| 327H03K | 20193 | 27.6 | Design of essential elements of an electric circuit (digital or analog) - clocks, amplifiers, pulse generators etc. |
| 438H01L | 20135 | 33.6 | Optimization of semiconductor manufacturing processes |
| 525C08L | 20116 | 11.7 | Polymer compositions employing combination of polymers as laminates, composites, coatings to improve stiffness, impact-resistance, insulation, anti-ageing, heat-resistance, water-proofing etc. |
| 430G03F | 19414 | 18.7 | Photo-sensitive and optically active materials to enable color applications in displays and fabrication of semiconductors by etching/photolithography |

The 50 domains, accounting for 39.4% of total patents, are seen to cover a very wide range of technologies. Highly patented domains exist across the known technological enterprise. The largest domains are represented by technologies related to materials, manufacturing, plastics, oil & gas, combustion engines, batteries, control systems, electrical systems, electronic displays, semiconductors, electronics, testing, software, telecommunications, medications, medical devices etc. The distribution of improvement rates for these technologies, shown in Figure 3, is similar to the distribution for the full set (see Figure 2) except that a smaller percentage of domains have an improvement rate less than 25%. The largest 50 domains contain a few very fast improving domains (9 in all) with an estimated improvement rate above 100% per annum. These are mostly related to software systems such as data management systems, distributed computing, memory management etc. and telecommunications such as routing, switching, mobility management etc. There are also slow improving domains improving at an estimated improvement rate below 10% per annum. These are mostly related to pharmaceuticals, packaging, motors, oil and gas well drilling etc.

We also plot the improvement rates for the biggest 50 domains in Figure 4. Readers familiar with past work in technology studies will notice that this figure closely resembles Figure 1b from Basnet and Magee (2016) which shows a plot of annual rate of performance improvement for 28 technological domains based on empirical date from Magee et al. (2016)

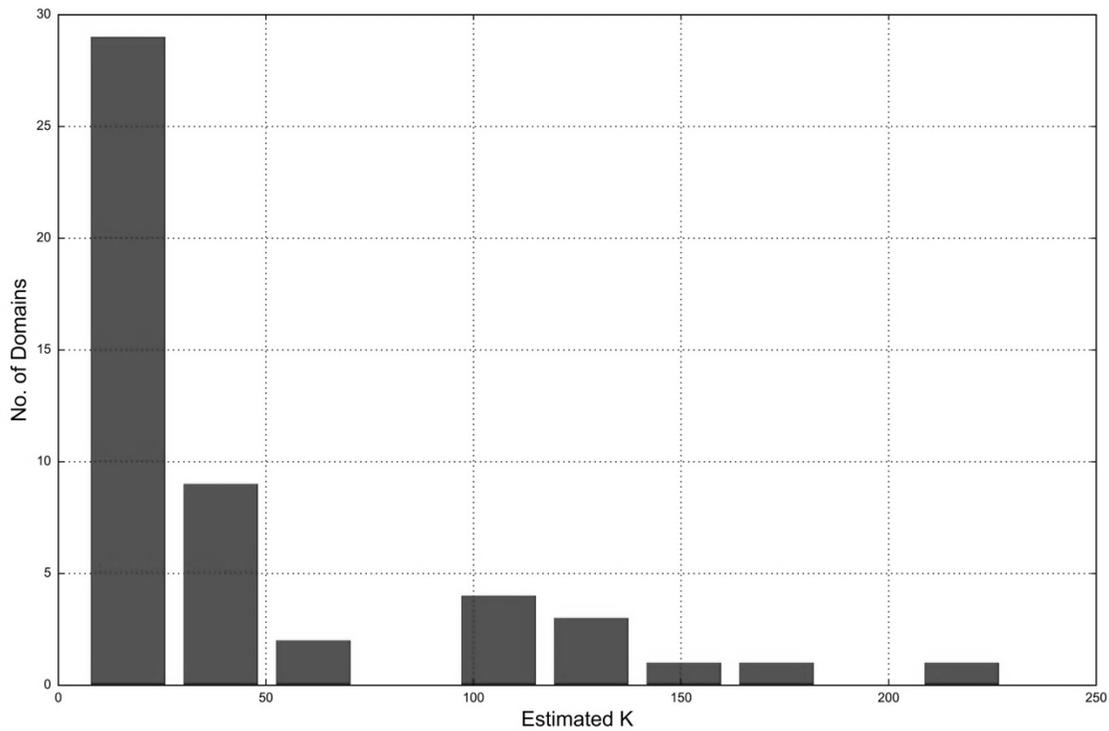

*Figure 3 Distribution of Estimated Improvement rates (K-values) for biggest 50 domains are similar to the distribution for the full set*

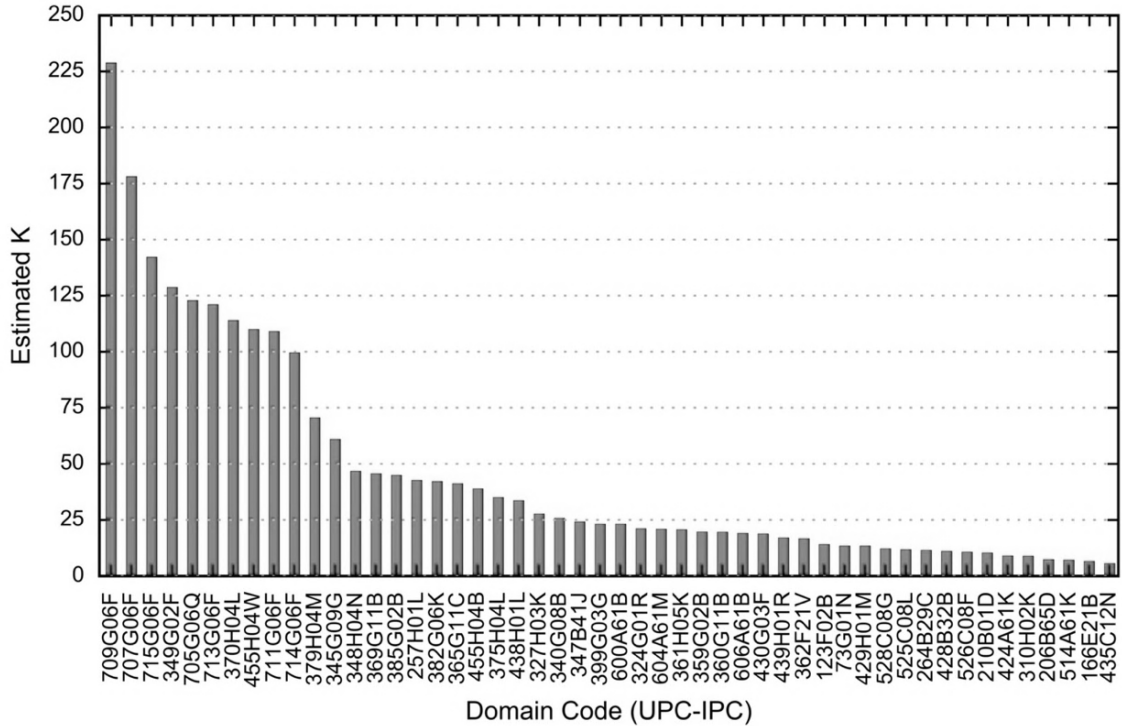

*Figure 4 Plot of Estimated Improvement rates (Estimated K-values) for biggest 50 domains are similar to the plot for empirical improvement rates for 28 domains from Basnet and Magee (2016) Figure 1b*

The diversity can also be observed using somewhat aggregated and more abstract classifications such as the 1-digit National Bureau of Economic Research (NBER) category of the corresponding UPC class of the technology domains and the IPC section of the corresponding IPC class of the technology domains.

The NBER classification was proposed by Hall, Jaffe, and Trajtenberg (2001) to link U.S. Patent Classification (UPC) classes to economically relevant technology categories. We mapped our domains to the NBER classification by matching the corresponding UPC code of the domain to the NBER category using "Class, technological category, and technological subcategory crosswalk" data from Hall et al. (2001). We were able to map 1731 of the 1757 domains to their NBER categories. We find that our set of technology domains is distributed broadly among all NBER categories with the least number of domains from Drugs & Medical and Computer & Communication categories, as shown in Figure 5.

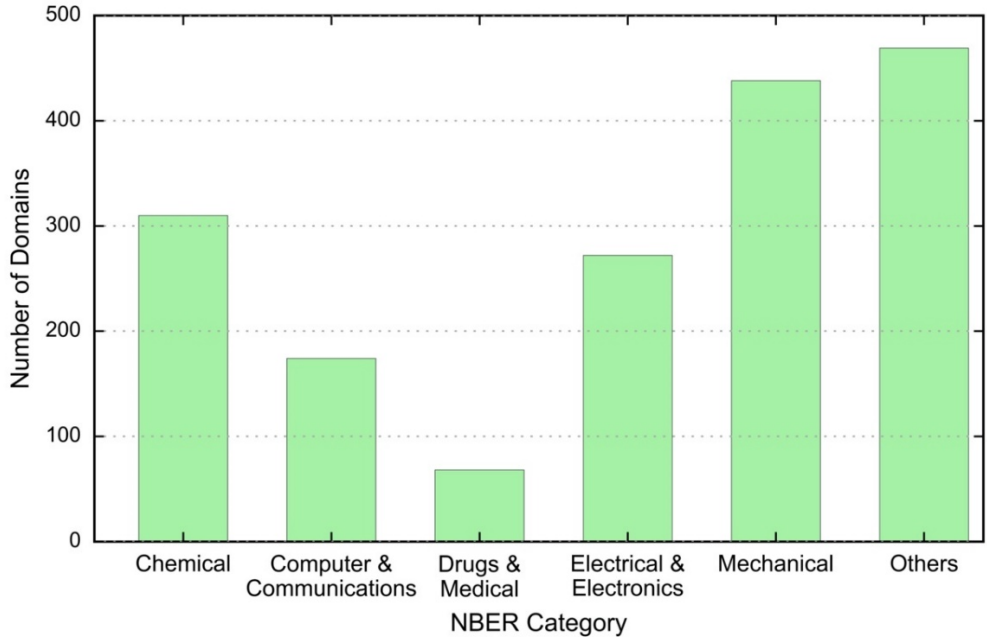

*Figure 5 Distribution of Technological Domains among NBER Categories*

We also investigate the distribution of domains by IPC sections. We mapped our domains to IPC sections by extracting the first digit of the corresponding IPC code of the domain. The one-digit code also refers to the section of the class. We again observe a broad distribution of domains across all sections. There are fewer domains belonging to Textiles; Paper and Fixed Constructions section, as shown in Figure 6.

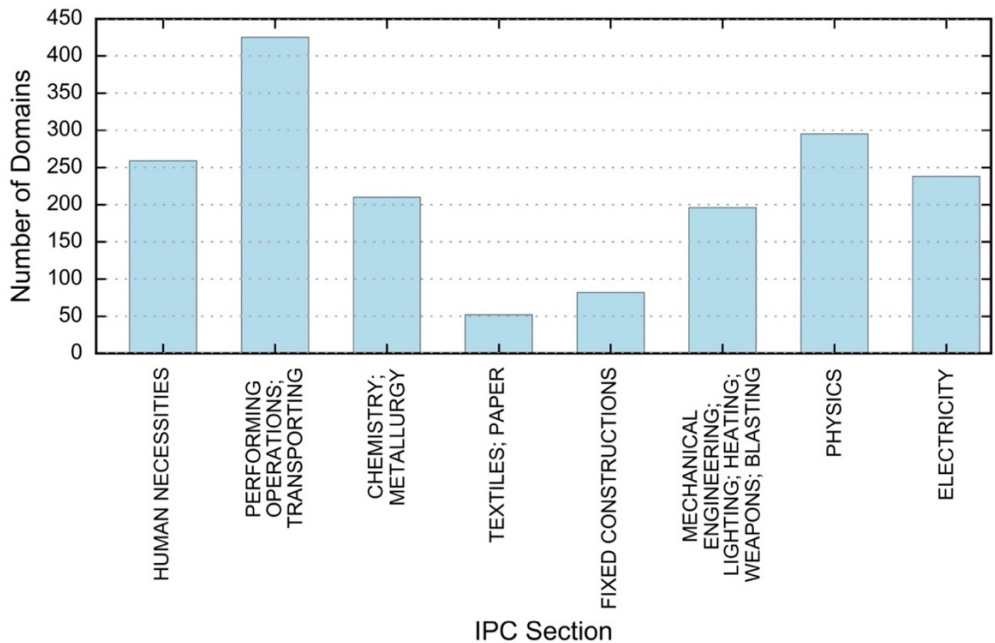

*Figure 6 Distribution of Technological Domains among IPC Sections*

**5.2 Software domains are generally improving more rapidly than other types of domains**

While most domains are indeed slow, we also find some very fast domains in the dataset. Table 7 identifies the 20 fastest improving domains. In contrast to the wide variation of technologies seen in Table 6, Table 7 shows much less diversity among technologies with most encompassing software improvements as an essential part of the domain.

**Table 7- Fastest improving 20 domains**

| Domain Code (UPC-IPC) | Domain Size | Estimated K | Domain Description |
|---|---|---|---|
| 709G06F | 64666 | 228.8 | Network management specifically client-server applications |
| 719G06F | 1834 | 213.9 | Dynamic information exchange and support systems integrating multiple channels |
| 726H04L | 1870 | 202.5 | Securing Enterprise Networks by system architecture (including security policies), user authentication, on the enterprise network, VPNs and defense mechanisms against DDoS attacks |
| 709G06Q | 313 | 196.4 | Network messaging system including advertisement |
| 726G06F | 6870 | 195.9 | Enterprise networks access management by individual users |
| 709H04L | 4779 | 194.3 | Network address and access management |
| 713H04L | 7436 | 193.4 | Data Encryption systems, including hybrid software/hardware systems and protocols for encryption, security associated with access and other security issues |
| 725H04N | 5581 | 193.2 | Content delivery in video distribution systems |
| 707G06Q | 328 | 192.8 | Automated data collection and information dissemination especially for ecommerce |
| 725G06F | 100 | 185.7 | Information delivery in video distribution systems |
| 715G06N | 123 | 182.3 | Information presentation methods on the web |
| 715G06Q | 149 | 179.3 | Business process automation |
| 707G06F | 53644 | 178.1 | Data management (including databases and novel data structures) for enabling and automating ecommerce activities |
| 73F02D | 287 | 175.9 | Software for detecting, measuring, estimating and calculating parameters for IC engine control |
| 380H04N | 244 | 174.4 | methods and apparatus for mixing encrypted digital data with unencrypted digital data |
| 717G06F | 12388 | 165.7 | Software delivery methods over internet including installation, testing, updates, packaged applications, web containers etc. and isolation of programs and threads in the processing environment |
| 718G06F | 3200 | 160.7 | Dynamic management of tasks and service requests on computing systems specially web servers by scheduling and virtualization |
| 725H04H | 178 | 154.9 | Systems for collecting user information in broadcast system |
| 705G07B | 671 | 147.2 | IT based Prepayment for services |
| 715G06F | 20979 | 142.1 | Compilation of related content/actions into a user-friendly graphic user interface based on context |

The fastest improving 20 domains are all software related. As shown in Table-7, technologies relating to the internet in general and enterprise network management in particular are estimated to be the fastest improving technologies in our set. These include technologies for effectively managing networks, network security, personalization and delivery of content over internet, software delivery over networks and business process automation, among others.

To date most estimates of software improvement rates have tended to focus mostly on the complexity related performance of well-known algorithms and some of these have indicated very high rates of improvement (Bentley, 1984; Richards and Shaw, 2004; Reed et al., 2005; Grace, 2013; Leland, 2016) along with some slower rates. Here, we observe that software systems are estimated to be improving even faster than observed rates for many algorithms.

As mentioned above only 201 of the 1757 domains, that is less than 11.5% of all domains, are improving at more than 36.5% per year (the estimated rate for integrated chips pertaining to Moore's law (Benson et al., 2018)). We call this set of domains as Moore201. A large number of these are software and telecommunications related. We observe the distribution of Moore201 domains again using the NBER classification and the IPC section.

Figure 7 shows the distribution of Moore201 domains by the NBER category. Similar to the top 20 domains, an overwhelming majority of these domains belong to the Computer & Communications category. Recall from Figure 5, that the Computer & Communication category accounts for fewer domains in the overall distribution by NBER category. 174 domains i.e. only 10% of the total domains are in Computer & Communication category. 106 of these 174 are in the Moore201 set.

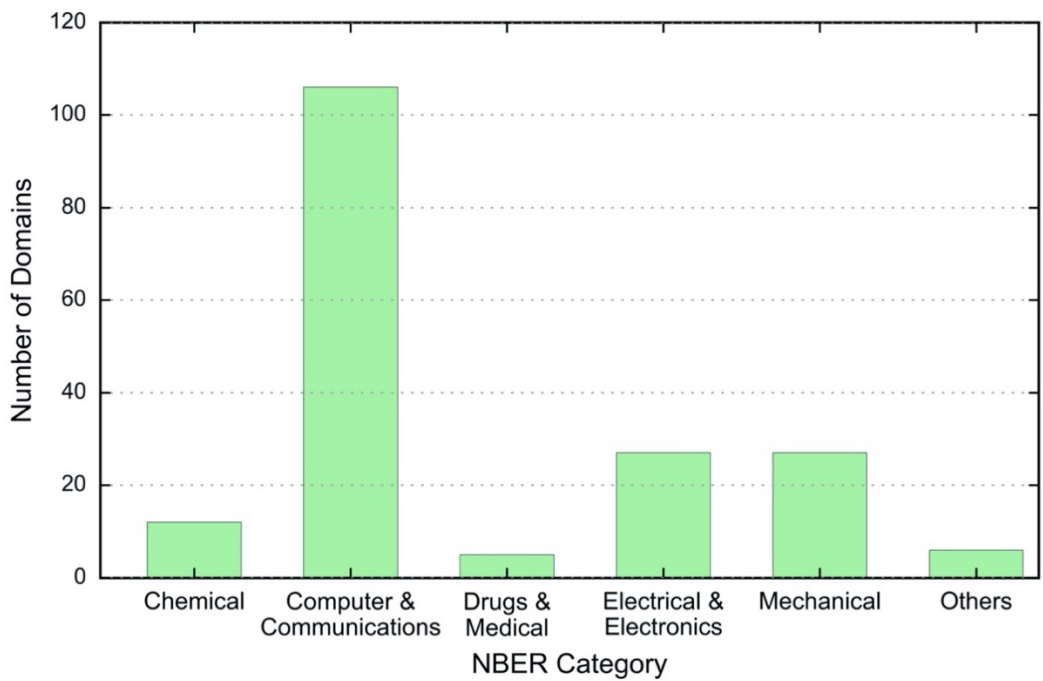

*Figure 7 Distribution of Moore201 among NBER categories*

Figure 8 shows the distribution of Moore201 domains by the IPC section. The share of Moore201 is highest for domains with corresponding IPC classes belonging to IPC sections G and H. A closer inspection of these domains, shows that 99 of these 201 domains correspond either to class G06 (Computing; Calculating or Counting) or class H04 (Electric Communication Technique). As seen in Table 7, the fastest improving 20 software domains also correspond to IPC classes G06 (software related) as well as H04 (telecommunication related).

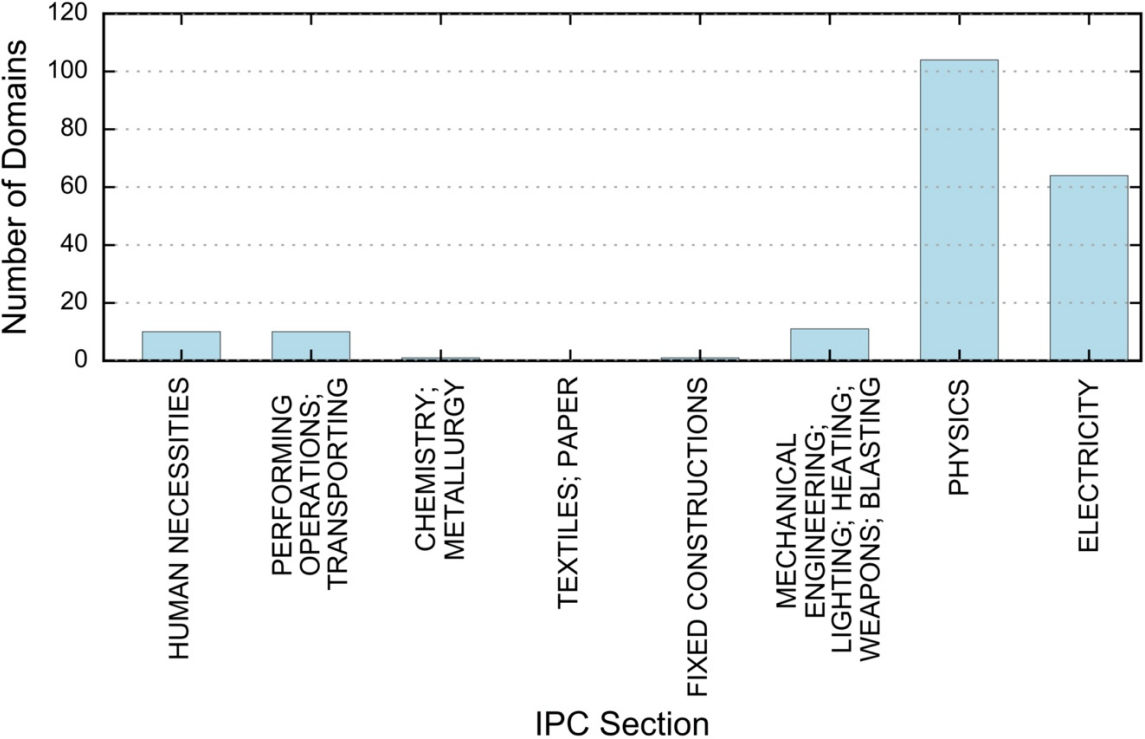

*Figure 8 Distribution of Moore201 among IPC Sections*

**5.3 Technologies related to simple mechanisms improve slowest**

Table 8 shows the slowest improving 20 domains. There is again little diversity among the slowest domains. The slowest domains are mostly related to simple mechanisms such as hitches, gauges, batons, hangers, cowls, hammocks etc.

**Table 8 Slowest 20**

| Domain Code (UPC-IPC) | Domain Size | Estimated K | Domain Description |
|---|---|---|---|
| 606A45D | 118 | 1.9 | Mechanical Skin treatment- Hair Removal and wrinkles |
| 431C11C | 124 | 2.8 | Candles |
| 7B25F | 330 | 2.8 | Handheld tools for cutting, scraping, drilling, punching etc. |
| 463F41B | 139 | 2.8 | Non-explosive weapons- Batons, tasers etc. |
| 16B25G | 351 | 2.9 | Handles |
| 280B60D | 1813 | 2.9 | Hitching assemblies for towing vehicles |
| 280B62H | 231 | 2.9 | Supports for two wheeled vehicle and locks |
| 223A41D | 192 | 2.9 | Physical manipulation of clothes, particularly hangers |
| 211A47L | 134 | 3.0 | Racks and Trays for household items |
| 33E04F | 114 | 3.1 | Gauges in construction work |
| 134B60S | 107 | 3.1 | Automatic vehicle washing |
| 227B25C | 2359 | 3.2 | Fastener driving apparatus- Power assisted nail guns, staplers etc. |
| 24A45F | 94 | 3.2 | Clips, hooks, straps, ties etc. for holding household items |
| 296B62J | 179 | 3.2 | Windshields, cowls etc. to protect the rider and reduce air resistance |
| 410B61D | 191 | 3.3 | Locking mechanisms to secure loads in transit |
| 362B25B | 129 | 3.3 | Integrated lamp in hand tools |
| 280B60S | 271 | 3.4 | Vehicle stabilization and miscellaneous operations |
| 5A45F | 115 | 3.4 | Hammocks and other miscellaneous camping equipment |
| 252C10M | 243 | 3.4 | Lubricants |
| 405E03F | 102 | 3.4 | Sewage/stormwater drainage methods and devices |

While 82.7% of the technological domains are improving at a rate of less than 25% per annum, comparatively few (only 134) are improving at a rate of less than 5% per annum. We call this Slowest134.

Figure 9 shows the distribution of Slowest134 domains by the NBER category. Most of these domains belong to the Mechanical and others category which is also reflected in Table 8. Figure 10 shows the distribution of Slowest134 domains by the IPC Section. The share of Slowest134 is highest for domains with corresponding IPC classes belonging to IPC sections A and B. A closer inspection of these domains, shows that 32 of these 134 domains correspond either to class A47 (furniture; domestic articles or appliances; coffee mills; spice mills; suction cleaners in general) or class B25 (hand tools; portable power-driven tools; handles for hand implements; workshop equipment; manipulators).

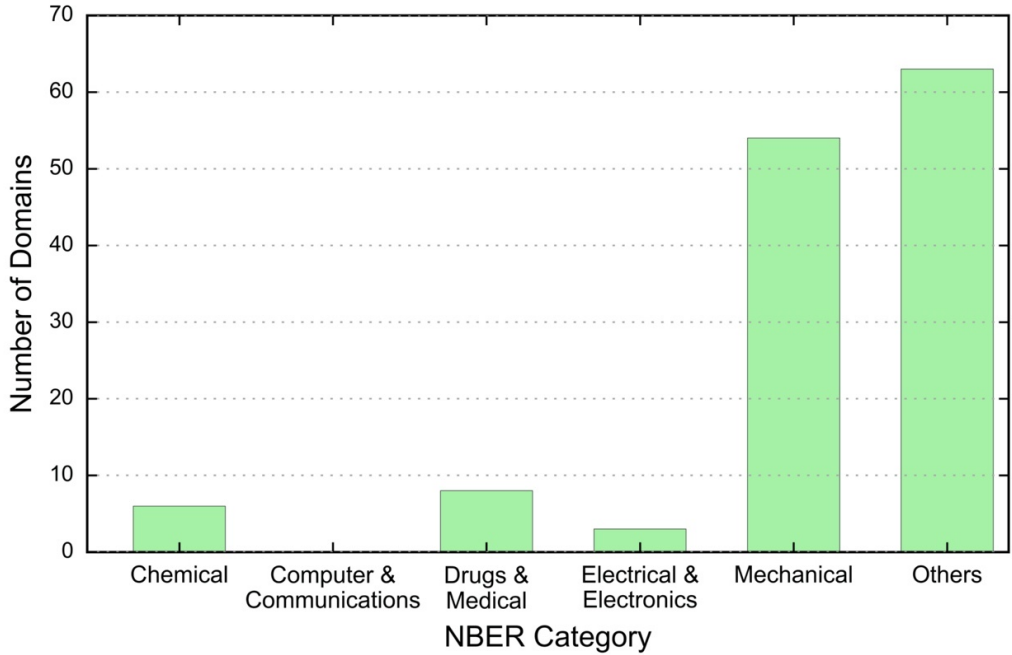

*Figure 9 Distribution of Slowest134 among NBER categories*

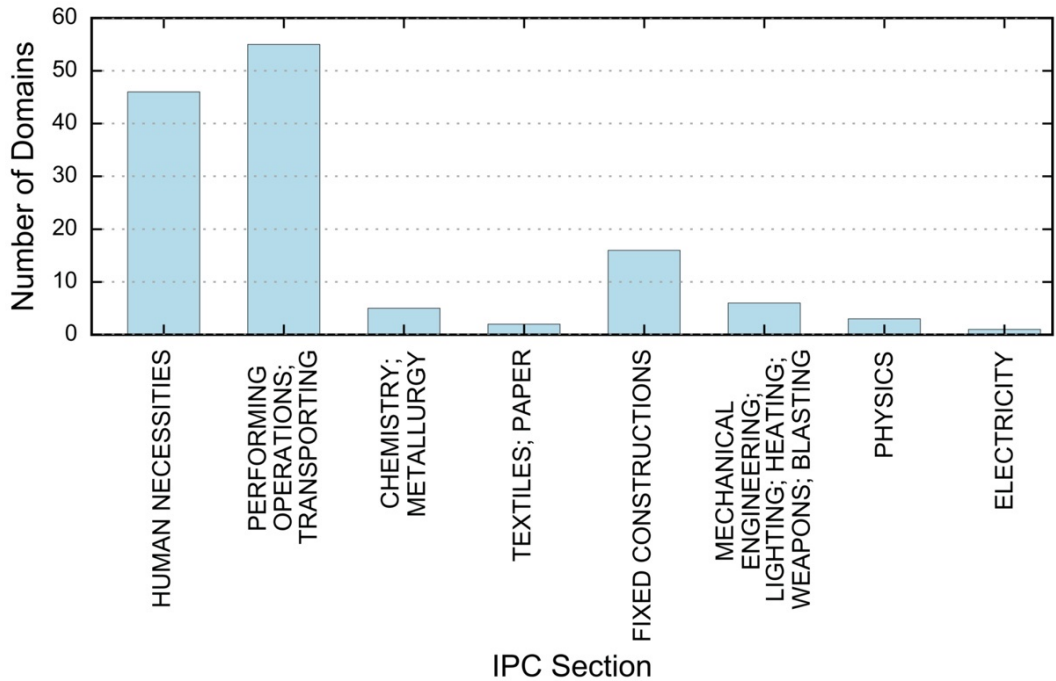

*Figure 10 Distribution of Slowest134 among IPC Sections*

**5.4 Very little correlation between Estimated K-values and size of domain**

Having a comprehensive list of estimated technology performance improvement rates and patent set for 1757 technology domains, we can investigate an additional important research question. Is there a relationship between the size of domains (measured by the number of patents- a proxy for R&D effort and investment) and the yearly improvement rate?

The method that we used to normalize centrality at the patent level control for the number of patents in the primary USPC class assigned to the patent. Therefore, at the patent level, by construction there is no effect of class size on patent centrality. However, in this work we are regrouping patents in technology domains, using a method that is only partially dependent on primary classification (we use the complete information on all classes assigned to a patent) and also incorporate information on IPC classes too. Therefore, it is possible that, if there would be a relationship between the technology domain size and the improvement rate, it would reveal itself in our estimates. If that would be the case, we could conclude that domains that are characterized by a large number of patents also experience faster improvement rates, possibly as a result of larger R&D investments in the technology (Messner, 1997; McDonald and Schrattenholzer, 2001; Schaeffer et al., 2004). However, we find that this is not the case for the technological domains we have investigated. In fact, we obtain some of the highest k values for small domains (less than 500 in size) and relatively low k values can also be seen at large domain sizes. We also observe some very-fast big domains and very-slow small domains. While the size of domain is a significant independent variable (see Table 9), the low r-square value of 0.0224 indicates that little of the variation in k can be explained by domain size variation (the R&D "effort").

| Intercept | 0.1791 |
|---|---|
| | (0.006) |
| Size of Domain | 4.544e-06*** |
| | (6.86e-07) |
| | |
| R-squared | 0.024 |
| No. of observations | 1757 |

*Table 9 Results from simple OLS regression of estimated K vs. size of Domain*

## 6. Discussion and Conclusion

The results represent the first attempt at a complete yet granular survey of technological performance improvement rate across the entire spectrum of technology. Our survey of estimated improvement rates, the online technology search system and the analysis of the distribution of improvement rates across all technologies domains have important managerial and policy implications, especially for allocation of resources among competing priorities. Nevertheless, before discussing and interpreting the significant results, we first note some limitations concerning the work. First, we have not determined every improvement rate of possible interest: the domains with less than 100 patents (~10,000 domains) may contain some important emerging technologies; we have not even attempted to name all 1757 domains that we separately estimated rates of improvement for; perhaps most importantly prior work using COM (Benson and Magee, 2016; Guo et al., 2016; Benson et al., 2018; You and Park, 2018) has often found specific technologies to be more closely identified at deeper sub-groups within the UPC and IPC classes than the high- level classes we applied systematically in this work. As shown in the identification work, we have done, the higher level we used leads to coherent domains and recognizable technologies but future work could pursue deeper level sub-domains. A second possible concern is the method we used to eliminate doubly (or triply etc.) listed patents where we simply assigned all duplicate patents to only the largest domain in which they are found. A different approach is to start by combining domains with high overlaps and use the remaining overlaps to give a measure of interactive structure. In research we are pursuing now, we are following this approach and comparing interactions identified by overlap to those found from patent citations between domains. Importantly, neither the patent overlap nor possible missing sub-domains is likely to significantly upset our conclusions in this paper. This is because the prior work looking at sub-domains has generally found less variation in rates of improvement among sub-domains within a domain (Benson et al., 2018; Sharifzadeh et al., 2019) than among higher-level domains; therefore, this limitation is not likely to affect the improvement rate variation we identified herein. Similarly, when we compare improvement rates in specific domains before and after elimination of duplicative patents, we find no large changes in k.

This work indicates (despite the concerns just discussed) that the rate of improvement in performance for a technologically-comprehensive set of 1757 technologies varies from 1.5% to 228.8% per year. The technologies identified indicate that the fastest improving technologies (greater than 36.5% improvement per year) are almost all centrally dependent upon software. However, there is apparently greater diversity in the slowest improving (< 25%) domains indicating that there are many paths to slower improvement but limited ones to faster improvement rates. The quantitative theory explaining variation in improvement rates is consistent with this finding since low interactions among components (McNerney et al., 2011) is a major factor in improving performance improvement rates (Basnet and Magee, 2016). High modularity (low interactions) is a core part of modern software design practice (Baldwin and Clark, 1997; Baldwin et al., 2000) and may represent the "limited path" to faster improvement rates but there are many ways to have low modularity so many technologies continue to improve slowly.

Although much recent thinking about technological improvement ascribes higher R&D effort and investment as causing faster improvement (Messner, 1997; McDonald and Schrattenholzer, 2001; Schaeffer et al., 2004), our results showing lack of correlation between number of patents in a domain and annual performance improvement in the domain do not support this assertion. We do not consider this result surprising since several past results (Sinclair et al., 2000; Sagar and van der Zwaan, 2006; Nordhaus, 2014; Funk and Magee, 2015) have pointed out problems with cumulative experience-based models. It appears that that the over-riding importance of spillover from other technologies and from the total scientific knowledge pool dominate effort within a domain as being the important driver of improvement in any domain.

Knowledge of the distribution of improvement rates across all technologies domains, which is a key contribution of this work, allows knowing how often we should expect rapid performance improvements. The fact that, as we showed in the paper, most of the rapid improving technologies are software-based, suggests that, if investors, firms, or countries would like to boost productivity gains, they should target higher investments in these areas.

Our survey of estimated improvement rates, which we make available to the public, can be very informative to determine optimal portfolios of investments in technology. As suggested by Way et al. (2019) investments in multiple technologies face a trade-off between concentrating investments in one project to spur rapid progress as opposed to diversifying over many projects to hedge against failure. In this case, information on improvement rates is crucial to be able to compute optimal allocation. The online technology search system which serves as a more precise complement to the broad technology system survey presented above, can be a valuable aid in this regard. It enables technology managers and policymakers to quickly look up estimates of improvement rates for specific technologies (or domains) or groups of related technologies. We believe and hope that this will bring greater accuracy, precision and repeatability to the as yet fuzzy art of technology forecasting and by accepting feedback continue to develop a deeper connection between popular technology terminology and the 1757 domains described here.

**Appendix A: Effect of deduplication on estimated rate of improvement**

As explained in section 4.1, we deduplicated the patent datasets for each domain for simplicity. We also estimated improvement rates using the original patent datasets before deduplication (arising directly from overlaps). In this section we examine the distribution of the absolute and percentage difference between estimated rates of improvement (estimated K) from original dataset and deduplicated dataset. The estimated rates of improvement (estimated K) are reported as percentage change per annum.

|  | ORIGINAL DATASET ESTIMATED K (O) | DEDUPLICATED DATASET ESTIMATED K (D) | ABSOLUTE DIFFERENCE (O – D) | PERCENTAGE DIFFERENCE (100 X [O-D]/D) |
|---|---|---|---|---|
| **MEAN** | 19.65 | 19.19 | 0.46 | 3.49 |
| **STD** | 27.48 | 26.25 | 7.09 | 25.56 |
| **MIN** | 2.67 | 1.88 | -62.47 | -66.11 |
| **MAX** | 224.24 | 228.79 | 76.40 | 274.78 |

*Table A.1 Summary statistics for estimated rates of improvement (estimated K) from original dataset and deduplicated dataset.*

As seen in Table A.1, the mean of both the percentage difference and the absolute difference are quite small indeed. The percentage difference mean value of 3.49% shows that there is no large bias in the deduplicated dataset. While there are a few big outliers, the standard deviation of 25.56 implies that more than two-thirds of the values have a difference of less than 25.56 from the mean. The mean of absolute difference is 0.46 percentage points with a standard deviation of 7.09 which is quite small, suggesting that almost all difference values lie close to 0. This can also be seen clearly in the plot of the distribution of absolute difference in Figure A.1.

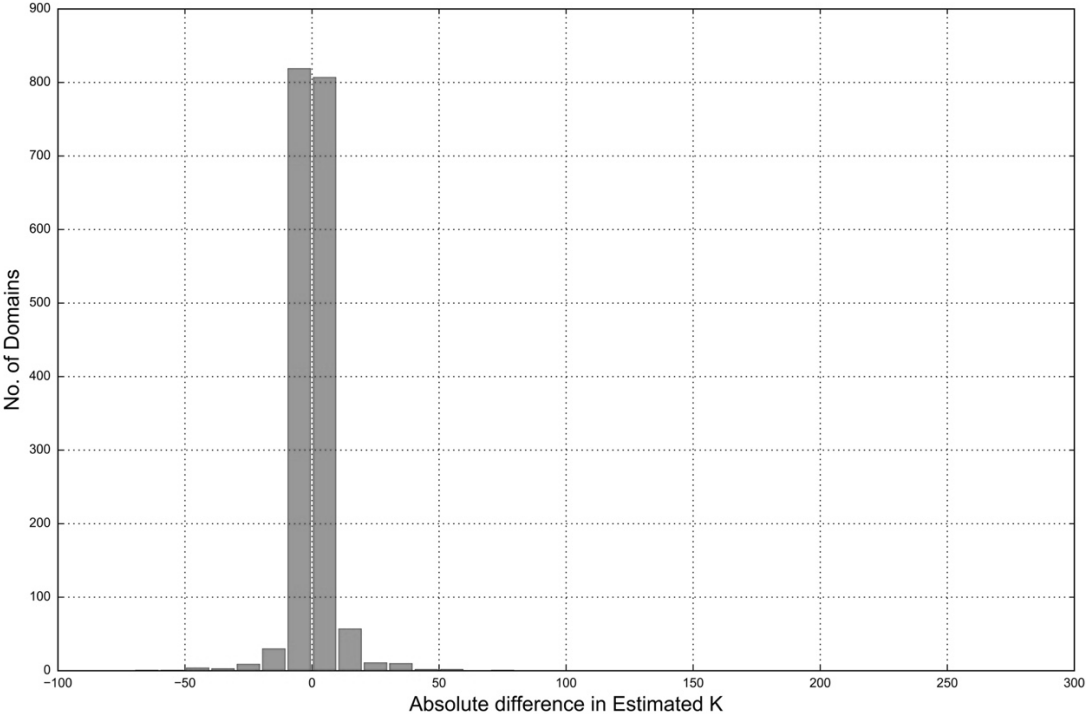

*Figure A.1 Distribution of absolute difference in estimated K*

**Appendix B: Normalization of patent centrality by citation network randomization**

Similar to what happens for scientific articles, patent writing and citations practices changes over time and across disciplines. This happens because of changes in patenting law and because sometimes, in some discipline it becomes customary to adopt a certain writing or citing practice (such as recurrently citing a given set of patents, to establish patentability of the subject matter). In some areas of technology development, there is also a stronger tendency of applicants to self-cite their previous patents or patents in the same technology subfield than in others. Furthermore, beside these "social biases" in citation practices, there are also some additional distortions that inflate or reduce the probability of being cited a certain number of times or make a certain number of backward citations for a given patent, everything else being equal. For instance, recent patents have less time to accumulate citations than older patents. Similarly, patents in recently emerged domains have less possible citing sources and citable targets than patents in older domains. All these differences make it difficult to compare centrality of different patents and to differentiate the signal of centrality from the effects of other factors, like the age of the patents, the number of citations made and received and the technological class(es) in which it appeared. For this reason, Triulzi et al. (2018) developed a method to separate the signal of centrality from these others confounding factors, by randomizing the overall citation network a thousand times and compute the centrality indicator for each patent in each of the 1000 randomized networks.

The randomization procedure consists in randomly swapping citations between pairs of patents under a series of constraints. Suppose that we observe in reality that patent A cites patent B and patent C cites patent D. These two citations are swappable (i.e. A would cite D and C cites B) in the randomized version of the citation network if A has the same grant year of C and B the same grant year of D and if one of the two following conditions apply:

- If A and B were assigned by the patent office to the same main technological classification, this classification must be the same of the one in which C and D are assigned
- If A and B were assigned to different classifications, A must have the same class of C.

These conditions ensure that in each version of the 1000 randomized networks, each patent preserve the same number of citations made and received, the same age profile of its citations made and received and the same share of citations made that go to patents classified in the same class. This automatically ensures that each technology class will have the same number of patents, the same number of citations made and received, the same distribution of citations made and received across different patent ages and the same share of citations falling within class and between classes in each of the 1000 randomized controls and in the observed reality. This allows computing a distribution of the centrality indicator of each patent (and of the average centrality of any given group of patents) across a thousand random but plausible worlds that resemble reality in each key characteristic but the one we are studying. We can then express the strength of the centrality signal for each patent as a z-score of the observed centrality value given the mean and the standard deviation of this distribution. We further normalize the z-score in a space from zero to one by taking the rank percentile of the z-score for patents granted in the same year. This, as explained in Triulzi et al. (2018), takes care of another possible source of biases, which is the empirical fact that the range of possible z-scores is a function of the indegree and outdegree of a patent, which in turns is a function of the year in which they are granted.

In this paper, we then compute the average value of the normalized centrality of a patent in a technology domain computed three years after they are granted. This is the predictor that we plug in Equation 2 (see Section 3.2) to calculate the estimated improvement rate.

**Appendix C: Normality tests and best fit of the distribution of mean patent centrality across the 1757 technology domains.**

We performed a series of normality tests to examine the possibility that the distribution of centrality across domains could reflect a random sampling of patents from the overall population. Figure C.1 shows the distribution of the mean centrality (calculated three years after the patent is granted) for all 1757 technology domains. The distribution is overlaid by the best fitted probability density function (an exponentially modified Gaussian distribution, with the rate, location and scale parameters respectively $1/2.057$, $0.313$ and $0.0586$[16]) and two comparison distribution, the Gaussian and the log-normal one. The sum of the squared errors for these three distributions is 7.67 (with an AIC of 120.6 and BIC of -9524.8) for the exponentially modified Gaussian, 11.88 (with an AIC of 147.6 and BIC of -8755.17) for the log-normal and 41.36 (with an AIC of 106.7 and BIC of -6571.9) for the Gaussian. This clearly shows that the distribution is not normal, nor is log-normal.

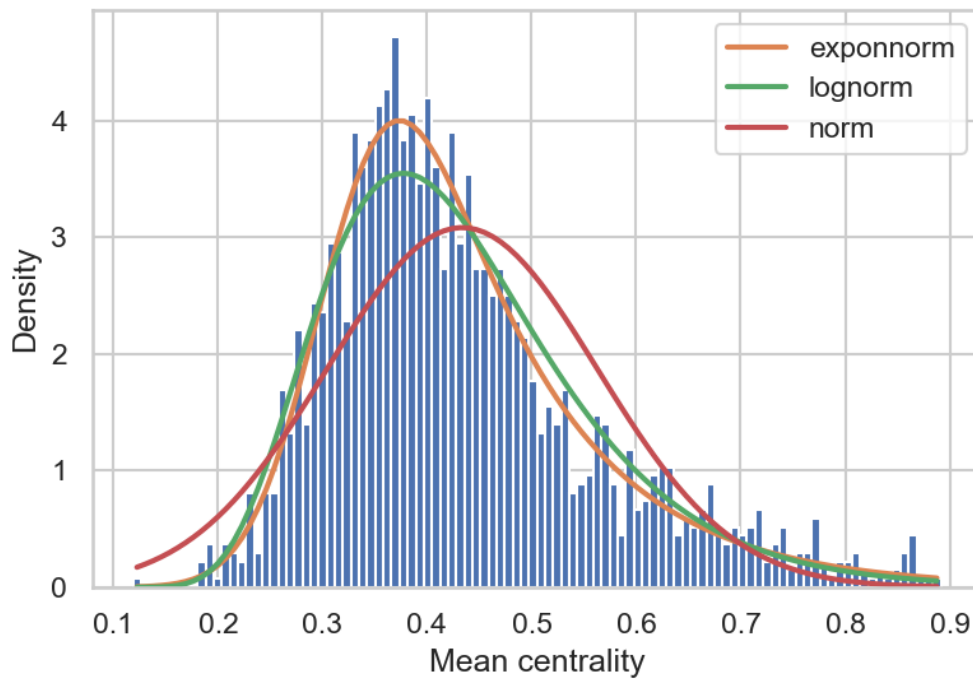

*Figure C.1: Fitting the distribution of mean centrality across the 1757 domains*

---

[16] The parameters have been obtained by fitting the distribution with the package Fitter in Python (https://fitter.readthedocs.io/en/latest/). For more information on the exponentially modified Gaussian distribution and an explanation of its parameters see: https://docs.scipy.org/doc/scipy-1.2.1/reference/generated/scipy.stats.exponnorm.html and https://en.wikipedia.org/wiki/Exponentially_modified_Gaussian_distribution.

However, we also performed a series of normality tests, reported in <mark>Table </mark>C.1, which unequivocally reject normality.

| Test | Statistic | P-value |
|------|-----------|---------|
| Shapiro-Wilk | 0.937 | 0.000 |
| D'Agostino's Chi-squared | 240.553 | 0.000 |
| Anderson-Darling | 32.758 | 0.000 |
| Kolmogorov Smirnov | 240.553 | 0.000 |

*Table C.1: Normality test results for distribution of mean centrality across the 1757 domains*